\documentclass[aps,floats,superscriptaddress,twocolumn,pre]{revtex4}
\usepackage{amssymb,amsmath}
\usepackage{amsmath,amssymb}
\usepackage{graphicx}
\usepackage{psfrag,textcomp}

\def\beq{\begin{equation}}
\def\eeq{\end{equation}}
\def\bea{\begin{eqnarray}}
\def\eea{\end{eqnarray}}
\begin{document}
\title{Generic nonequilibrium steady states  in an exclusion process  on an inhomogeneous ring}
\author{Tirthankar Banerjee}\email{tirthankar.banerjee@saha.ac.in}
\author{Niladri Sarkar}\email{niladri.sarkar@saha.ac.in}
\author{Abhik Basu}\email{abhik.basu@saha.ac.in}
\affiliation{Condensed Matter Physics Division, Saha Institute of
Nuclear Physics, Calcutta 700064, India}
\date{\today}

\begin{abstract}
We consider a one-dimensional totally asymmetric exclusion process
on a ring with extended inhomogeneities, consisting of several
segments with different hopping rates. Depending upon the underlying
inhomogeneity configurations and for moderate densities, our model
displays both localised (LDW) and delocalised (DDW) domain walls and
delocalisation transitions of LDWs in the steady states. Our results
allow us to construct the possible steady state density profiles for
arbitrary number of segments with unequal hopping rates. We explore
the scaling properties of the fluctuations of LDWs and DDWs.

\end{abstract}

\maketitle

\section{Introduction}
 Totally asymmetric simple exclusion process (TASEP) in
one dimension (1D) serves as a paradigmatic model for systems out of
equilibrium, where fundamental questions concerning nonequilibrium
statistical mechanics can be asked and explicitly answered.
Phenomenologically, it is used as a simple physical description for
a variety of systems ranging from quasi-1D transport in cell biology
contexts, e.g., ribosome translocations along messenger RNA
(mRNA)~\cite{zia} to fluid motion along artificial crystalline
zeolitical structures~\cite{zeo} and movement of unidirectional
vehicular traffic along roads~\cite{traffic}. The novelty of TASEP
as an open system lies in its ability to display boundary-induced
phase transitions~\cite{tasep-rev}. Several different variants of
TASEP~\cite{lktasep,andrea}, including effects of inhomogeneities
have been studied~\cite{def1,def2}, shedding insight for
nonequilibrium physics.

In this article, we consider the steady state densities in TASEP in
an inhomogeneous system with periodic boundary conditions, e.g., a
closed ring. Unlike open systems, closed-ring systems admit strict
particle number conservation, which is known to affect the universal
scaling behavior near a critical point in uniform
systems~\cite{halp}. However, TASEP on a uniform ring cannot display
spatially non-trivial steady state density profiles, due to the
spatial translation invariance of the model. Only when translational
invariance is broken explicitly, e.g., by connections with diffusive
lanes, presence of defects or appearance of junctions and branches,
inhomogeneous density profiles may
appear~\cite{lebodadu,mustansir,erwin1}. In general, steady state
properties of asymmetric exclusion processes on a ring with
inhomogeneities have been studied by using various models. For
instance, Ref.~\cite{lebodadu} considered TASEP on a ring with one
point defect (i.e., a single slow site) and found LDWs in the steady
states under certain conditions. Subsequently, Ref.~\cite{mustansir}
extensively discussed disordered asymmetric exclusion processes by
using the {\em fully segregated model} (FSM), where the system has
two parts with different hopping rates; see also
Refs.~\cite{stinch1,nossan} for detailed studies on the various
aspects of the effects of quenched disorders on TASEP,
Ref.~\cite{stinch} discussed the effects of quenched disorders on
TASEP and their connections with the dynamic scaling of
Kardar-Parisi-Zhang~\cite{kpz} surfaces.

In the present work, we are generally motivated to study the
possible density profiles with finite number of segments having
unequal hopping rates as consequences of the mutual interplay
between conservation laws and the scale of inhomogeneities in closed
exclusion processes. We extend and complement the existing studies
on closed TASEP with inhomogeneities by considering a model of
exclusion process on a closed inhomogeneous ring having multiple
segments with unequal hopping rates.  Our model reveals
nonequilibrium phase transitions including delocalisation
transitions of LDWs into delocalised domain walls (DDWs), distinctly
different from boundary induced phase transitions in an open TASEP
(see, e.g., Ref.~\cite{krug} for boundary induced phase transitions
in driven diffusive systems).  We discuss how these may be
generalised for larger number of segments. In addition, we
numerically explore the nature of fluctuations of LDWs and DDWs in
the system and the associated delocalisation transition. Our results
should be of significance in understanding the phenomenologies of
ribosome translocations along closed mRNA loops ({\em circular
translation of polysomal mRNA})~\cite{chou,kopeina} with clusters of
{\em slow codons} along which ribosome translocations are
inhibited~\cite{robinson,sorensen,cluster}. We use analytical
mean-field theories (MFT) complemented by extensive Monte Carlo
simulation (MCS) studies for our work. Our main findings in
this article are as follows: (i) When there are three or four
segments of all unequal hopping rates, there should be one LDW in
the system with the slowest segment being in the maximal current
(MC) phase for moderate densities, (ii)  for moderate densities with
four segments having two non consecutive slowest segments (of equal hopping rates),
instead of a single LDW, there will be two DDWs, each formed behind one of the
slowest segments; the slowest segments themselves are in their MC
phases, (iii) we use our results from the three- and four-segment
models to infer about the possible general forms of the steady state
density profiles for arbitrary number of segments with different
hopping rates, and (iv) we numerically uncover the scaling
properties of widths of the LDW fluctuations, both near and away
from delocalisation transitions. The remaining parts of this paper
are organised as follows.In Sec.~\ref{two} we present a brief review
of the FSM. In Secs.~\ref{three} and \ref{four} we
construct our models with three (Model I) and four (Model II)
segments, and study in details the corresponding steady state
density profiles. In Sec.~\ref{general}, we use our results for
three and four segments to comment on systems with arbitrary number
of segments. Lastly, in Sec.~\ref{scaling}, we investigate the
scaling of the fluctuations of LDW and DDWs in our MCS studies. In
Sec.~\ref{summ}, we summarise and conclude.

\section{Review of the two-channel system}\label{two}

\begin{figure}[htb]
\includegraphics[height=5cm, width=7.6cm]{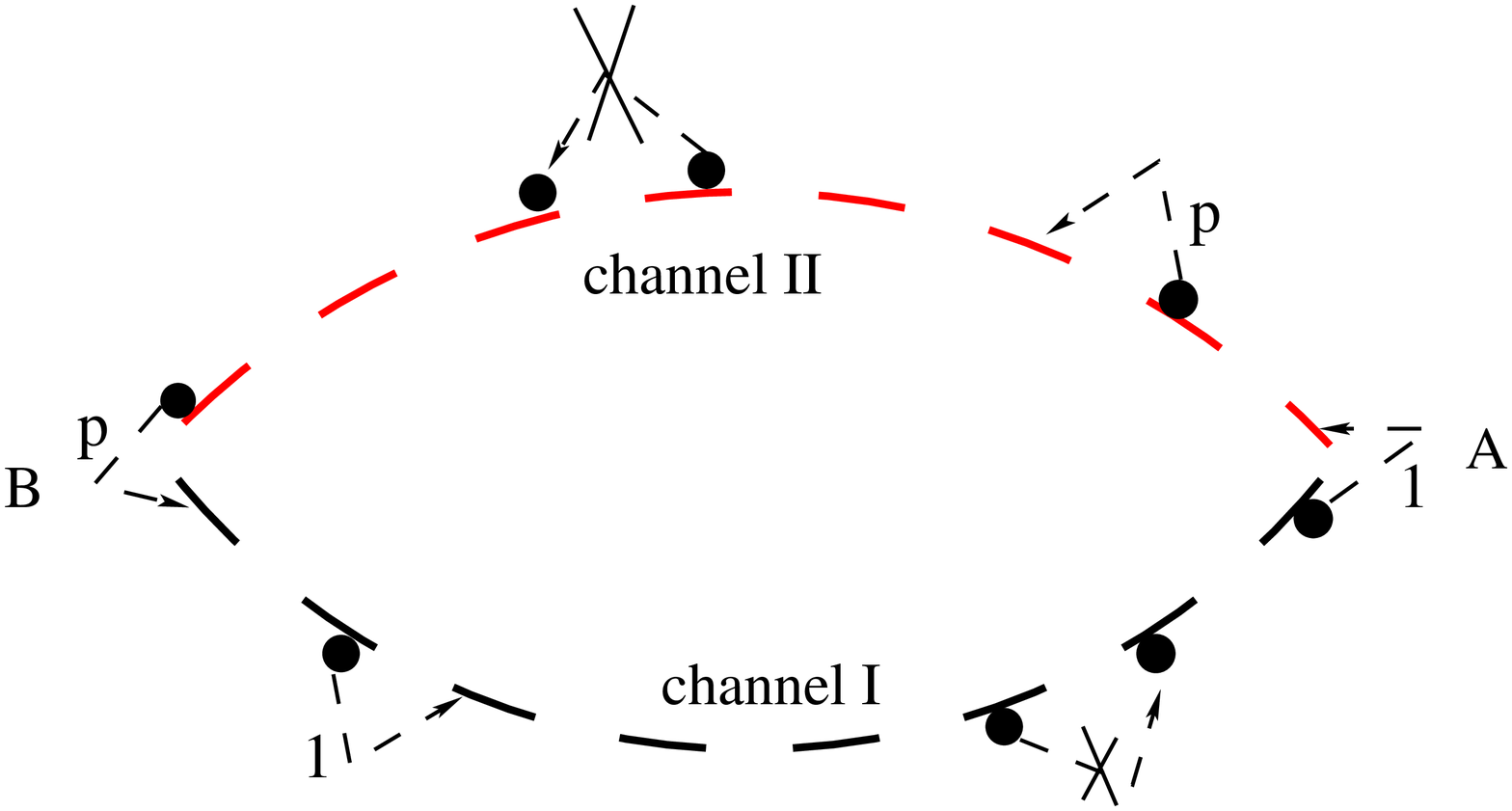}
\caption{Schematic diagram of the FSM model. CHI (black)
has a hopping rate 1, CHII (red) has a hopping rate of $p<1$ (see
text).} \label{model}
\end{figure}

In order to set up the background,
it is instructive to first review the results from the analogous
two-segment system (FSM), as elucidated in Ref.~\cite{mustansir}. It
has two parts, CHI and CHII, of hopping rate 1 and $p$,
respectively; see Fig.~\ref{model} for a schematic diagram. Assume
equal lengths for CHI and CHII. If we represent a phase
of the whole system by $(X,Y)$, then $(X,Y)$ can be any of low density (LD),
high density (HD) , domain wall or co-existence (DW) and maximal current (MC) phase
of CHI and CHII, respectively.
As discussed in  Ref.~\cite{mustansir} in great details,
the system can be only in three out of all possible sixteen phases. These
three phases are (LD,LD), (HD,HD) and (DW,MC).

 The above physical picture~\cite{mustansir} can be made quantitative as follows. Consider first the (LD,LD) or (HD,HD)
 phases { and ignore the boundary layers (BLs) (having vanishing width in thermodynamic limit) in MFT,}
  for which the bulk densities $n_I$ and $n_{II}$ are spatially constants in the steady states.
 Then, $J_I=J_{II}$ in the steady states gives
 \bea n_I(1-n_I)=pn_{II}(1-n_{II}).
\label{currentcon1} \eea With the particle number density 
given by \bea n=\frac{n_I +n_{II}}{2}, \eea we obtain \bea
n_{II}=\frac{-b \pm \sqrt{b^2-4ac}}{2a} \eea with $a=(p-1)/4$,
$b=n-\frac{(1+p)}{4}$ and $c=n/2-n^2$. These yield densities
$n_{II}$ and $n_I$ when both of them are in their LD or HD phases.
The two thresholds for $n$, namely, $n_L,n_U$, which mark the boundaries between
LD/MC or HD/MC phases for CHII (or, equivalently, the boundaries
between LD/DW or HD/DW phases for CHI) may be obtained by setting
$n_{II}=1/2$ (see below).

Consider  CHII to be in its MC phase, while CHI is to have a
coexistence of LD and HD phases in the form of a DW or a
discontinuity in its density profile. By using $J_I=J_{II}$, we get
\bea n_I^2-n_I+{p \over 4}=0. \eea This gives us the solutions
$n_I={1 \pm \sqrt{1-p} \over 2}$, both corresponding to the same
$J_I$.  The two solutions meet at a point $x_I$ in the form of a DW
in CHI. More specifically, since CHII has a hopping rate $p$, lower
than the hopping rate (unity) in CHI, on simple physical ground we
expect particles to pile up {\em behind} the junction B, so that
when there is a discontinuity in $n_I(x)$: 
  $n_I =(1-\sqrt{1-p})/2 <1/2$ starting from $x=0$, then jumps discontinuously at $x_I$ ($0<x_I<1-l$) to
  $n_I=(1+\sqrt{1-p})/2 >1/2$ up to $x=1-l$. Thus, in terms of
the standard notations of a TASEP with open boundaries, we identify
the particle injection rate $\alpha=(1-\sqrt{1-p})/2<1/2$ and
particle extraction rate $\beta=(1-\sqrt{1-p})/2<1/2$, thus giving
$\alpha=\beta<1/2$, in agreement with the condition for coexistence
in open TASEP. With a DW at $x_I$, we have
 \bea n_I=\alpha_I+\theta(x-x_I)(1-\alpha_I-\beta_I), \eea
$\theta(x-x_I)$ is the Heaviside step function. Hence, there is a
jump in density by a value $1-\beta_I-\alpha_I$ at $x=x_I$ giving
the height of the DW. Using particle number conservation
$(\int_0^{1-l}n_I
+\int_{1-l}^1 n_{II})dx = n$, we find
\bea x_I=-{n-({\sqrt{1-p} \over 2})({1 \over 2})-{1 \over 2} \over
\sqrt{1-p}}. \label{dw1p} \eea 
The phase boundaries (values for $n_L$ and $n_U$) are obtained by setting $x_I=0$ and $x_I=1/2$
(see Ref.~\cite{mustansir} for a phase diagram).

\section{Model I: Three segment System}\label{three}

\begin{figure}[htb]
\includegraphics[height=5cm]{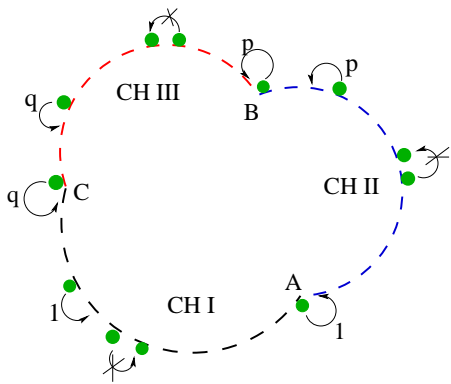}
\caption{Schematic diagram of the 3-sector model. CHI
(black) has a hopping rate 1, CHII (blue) has a hopping rate of
$p<1$  and
 CHIII (red) has a hopping rate $q<p$ (see
text).} \label{model1}
\end{figure}

We consider a closed 1D ring having total $N$ sites with three parts
of equal lengths but with {\em different hopping rates} (henceforth
Model I). Equivalently, our model can be viewed as a TASEP on a ring
with {\em  extended inhomogeneities {\rm or} defects}; see
Fig.~\ref{model1} for a schematic diagram. The three parts, marked
CHI, CHII and CHIII (in Fig.~\ref{model1}) are equal in length
having $N/3$ sites each. Segments CHI, CHII and CHIII, respectively
have hopping rates unity (with sites $i=1,...,N/3$),  $p<1$ (with
sites $i=N/3+1,2N/3$) and $q<p<1$ (with sites $i=2N/3+1,...,N$).
Particle exchanges between each pair of channels are exclusively
allowed at their junctions A,B and C with dynamic rules defined by
the originating site. If $n_{Ii}, n_{IIi}$ and $n_{IIIi}$ represent
the average number denisities in $i$-th sites of CHI, CHII and
CHIII, respectively and $N_{tot}$ denotes the total number of
particles in the system, we can write,
\begin{equation}
 (n_{I}+n_{II}+n_{III})\frac{N}{3}=N_{tot} ,
\end{equation}
where, $n_I=\sum_in_{Ii}$ (similarly for $n_{II}$ and $n_{III}$).
The mean number density for the total system, $n$ can thus be defined as
\begin{equation}
n=\frac{N_{tot}}{N}=\frac{n_{I}+n_{II}+n_{III}}{3}. \label{ratio1}
\end{equation}
We are, thus, left with a three-dimensional parameter space spanned
by $n,\,p$ and $q$, each of which is a bulk quantity and confined
between 0 and 1. Tuning the phase parameters, we obtain the various
possible phases of the system.

\subsection{MF analysis of Model I}\label{mf_three}

We use and extend the idea outlined above. In our MF analysis we treat CHI, CHII and CHIII as separate TASEP
lanes joined at junctions A, B and C, and analyse the phases of the
model in terms of the known phases of an open-boundary TASEP. For
convenience, we label the sites by a continuous variable $x$ in the
thermodynamic limit, defined by $x=i/N$, $0<x<1$. We represent a
phase of the whole system by $(X,Y,Z)$, where $X,Y,Z$ can be either
the LD, HD, MC or DW phases of CHI,CHII and CHIII, respectively.
Thus, all in all, there can be at most $4\times 4\times 4=64$ phases
in the whole system. We now discuss whether or not all the phases
are allowed by the constraints of the
system. 
With the bulk densities $n_I$, $n_{II}$ $n_{III}$ respectively, in
CHI, CHII and CHIII, the corresponding steady state currents are
given by $J_I=n_I(1-n_I)$, $J_{II}=pn_{II}(1-n_{II})$ and
$J_{III}=qn_{III}(1-n_{III})$, with $J_I=J_{II}=J_{III}$ in the
steady states.
\par
Consider the case where the particle motion is from
CHI to CHIII via CHII. We argue now that phases (LD,LD,HD),(LD,HD,HD) and
(LD,HD,LD) are ruled out. We use the following characterisation of the phases in a TASEP with open boundaries to delineate the phases in CHI, CHII and CHIII of our model:
\begin{itemize}
\item When a segment CHX (X= I, II, III) is in its low density (i.e.,$n_X < 0.5$), there is no BL in the vicinity of the entrance
of CHX. We denote the corresponding uniform steady state in the bulk by $\alpha_X$, which is same as the injection rate.
\item When a segment CHX (X= I, II, III) is in its high density (i.e.,$n_X > 0.5$), there is no BL in the vicinity of the exit end
of CHX. We denote the corresponding uniform steady state in the bulk by $1-\beta_X$, where $\beta_X$ is the corresponding extraction rate.
\end{itemize}

For concreteness, consider first the
possibility of the (LD,LD,HD) phase, with $\alpha_I$,$\alpha_{II}$
and $1-\beta_{III}$ being the bulk densities of CHI, CHII and CHIII,
respectively. 
Accordingly,  the
junction marked $C$ in Fig.~\ref{model1} has no BL,
where as the junction $B$ will have two BLs on both sides. We now use current conservation
across junction $C$ and in the bulk of CHI and CHIII. These yield
\begin{equation}
q(1-\beta_{III})\beta_{III}=q\beta_{III}(1-\alpha_{I}) =  \alpha_{I}
(1-\alpha_{I}).\label{nobl}
\end{equation}
Thus, we must have (i) $1-\beta_{III}=1-\alpha_I$ and (ii) $q\beta_{III}=\alpha_I$.
Therefore, the only consistent solution of the above is $q=1$, making
CHI and CHIII identical. Thus, if $q\neq 1$, there must be a BL at
$C$, which in turn rules out the (LD,LD,HD) phase. Similarly the
(LD,HD,HD) and (LD,HD,LD) phases also do not appear. Using the
particle-hole symmetry~\cite{tasep-rev}, the (HD,HD,LD),(HD,LD,LD)
and (HD,LD,HD) phases are also ruled out immediately. Furthermore,
CHI and CHII can never be in their respective MC phases
(corresponding to a currents $1/4$ and $p/4$, respectively), since, the maximum permissible
current in the system (with $q\neq 1,p$) is just $q/4<1/4, p/4$.
Hence, the system has no phases of the form (MC,$Y$,$Z$) or
($Y$,MC,$Z$) where $Y,Z$ can be any of LD, HD, MC or DW phases.
Additionally, we can make the following general observations
concerning the nature of allowed phases in the system. With
$J_I=J_{II}=J_{III}$ in the steady states,  for each of CHI, CHII
and CHIII in the LD (HD) phases, $n_I < (>) n_{II}< (>) n_{III}$.
For very low $n$, all the channels are to be in their respective LD
phases, with the bulk densities $n_{III}>n_{II}>n_I$. Thus, with
increasing $n$, $n_{III}$ reaches a value 1/2, at which CHIII enters
its MC phase, with $n_I$ and $n_{II}$ each still less than 1/2.
Using the particle-hole symmetry we draw the (equivalent) physical
picture of the borderline between the MC and HD phases of CHIII with
$n_I, n_{II}>1/2$. Hence, on physical grounds, we identify different
thresholds, $n_L$ and $n_U$, for $n$, that demarcate the LD and MC
phases, and MC and HD phases of CHIII, respectively. The values for
$n_L$ and $n_U$ are obtained below. Clearly, with $n_L<n<n_U$, CHIII
remains in its MC phase, i.e., $n_{III}=1/2$ (neglecting the BLs in
CHII) corresponding to a steady state current $q/4$ in the system.
Therefore, as $n$ rises from $n_L$ to $n_U$, all the added particles
must accumulate in CHI and CHII, such that $J_I=J_{II}=J_{III}=q/4$
is maintained.  This suggests a coexistence of LD and HD phases in
the form of a DW in CHI or CHII, so that with $n$ changing within
the interval $n_L<n<n_U$  the total particle content in CHI and CHII
changes due to shifts in the location of the LDW keeping $J_I$ and
$J_{II}$ unchanged.

With CHIII, having the lowest
hopping rate being in its MC phase with $n_3=1/2$, using the
continuity of current in each channel, we write
\begin{eqnarray}
 n_I = \frac{1 \pm \sqrt{1-q}}{2},\\ n_{II} = \frac{1 \pm \sqrt{1-q/p}}{2}.
 \end{eqnarray}
 The two solutions for each of $n_{I}$ and $n_{II}$ represent either HD or LD
 phase densities for CHI and CHII, respectively. Since on physical ground, particles start
 accumulating behind the slowest segment (CHIII here), depending on the exact configuration of the model,
 an LDW will be formed either in the segment with hopping rate 1 or with hopping rate
 $p$. Thus, given Fig.~\ref{model1}, an LDW forms in CHII and then
 shifts towards CHI as more and more particles are added.

Consider now an LDW in CHII. As per our arguments, CHI will be in
its LD phase given by $n_I=\frac{1 - \sqrt{1-q}}{2}$. In order to
maintain the same current $J_{II}$, $n_{II}$ will have two solutions
which will meet at a point say, $x_{II}^{w}$. Thus we have a
discontinuous jump in the density profile at $x_{II}^{w}\,
(1/3<x_{II}^{w}<2/3)$. As in the FSM~\cite{mustansir}, we identify
effective particle injection rate, $\alpha_2=\frac{1 -
\sqrt{1-q/p}}{2}<1/2$ and an effective extraction rate,
$\beta_2=\frac{1 - \sqrt{1-q/p}}{2}<1/2$, thus satisfying the
coexistence condition, $\alpha_2=\beta_2<1/2$.

Total particle number conservation in the model yields
\begin{equation}\label{3sector_x}
 \left(\int_0^{1/3}n_I + \int_{1/3}^{2/3}n_{II} + \int_{2/3}^{1}n_{III}\right)dx = n,
\end{equation}
The density profile, $n_{II}$ with an LDW at $x_{II}^w$ is given by
$\alpha_2 +\Theta (x-x_{II}^w) (1-\alpha_2 -\beta_2)$, where $\alpha_2
=\beta_2= \frac{1 - \sqrt{1-q/p}}{2}$. $\Theta$ is the Heaviside
step function. The jump in the density by the value
$1-\alpha_2-\beta_2$ at $x=x_{II}^w$, gives the height of the DW.

The position of the LDW can be evaluated from Eq. \ref{3sector_x},
which gives
\begin{equation}
 x_{II}^w = - \frac{n-1/2-\sqrt{1-q/p}/2+\sqrt{1-q}/6}{\sqrt{1-q/p}}.\label{3sector_DW_right}
\end{equation}
Similarly, if the DW is created in CHI, then CHII will be in its HD
phase and the corresponding position for the LDW will be
\begin{equation}
 x_{I}^w = -\frac{n-1/2-(\sqrt{1-q/p}+\sqrt{1-q})/6}{\sqrt{1-q}}. \label{3sector_DW_left}
\end{equation}
The phase boundaries are obtained by setting $x_I^w=0$ and $x_{II}^w
= 2/3$. For $x_I^w=0$, the equation for the boundary line reads
\begin{equation}
 n=n_U=1/2 + (\sqrt{1-q/p}+\sqrt{1-q})/6, \label{3sec_phb1}
\end{equation}
while for $x_{II}^w = 2/3$, we have
\begin{equation}
 n=n_L=1/2 - (\sqrt{1-q/p}+\sqrt{1-q})/6. \label{3sec_phb2}
\end{equation}
We observe that the phase boundaries do not change if the transport
direction was from CHI to Channel II via CHIII.

\begin{figure}[htb]
\includegraphics[height=5cm]{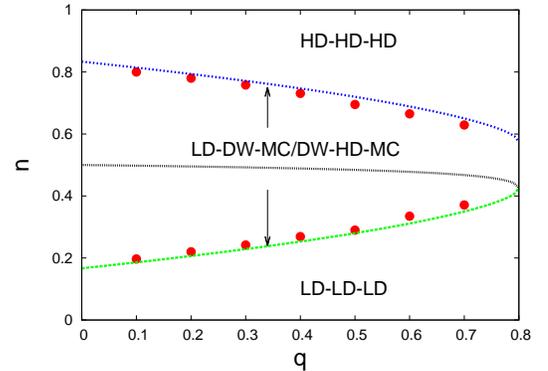}
\caption{Phase diagram of the 3-Channel system. Phase parameters
are $n$ and $q$, with $p=0.8$. The black line represents the boundary
when the LDW is formed at the junction of CHI and CHII. Points represent
MCS results.} \label{3_phase}
\end{figure}

\begin{figure}[htb]
\includegraphics[height=5cm]{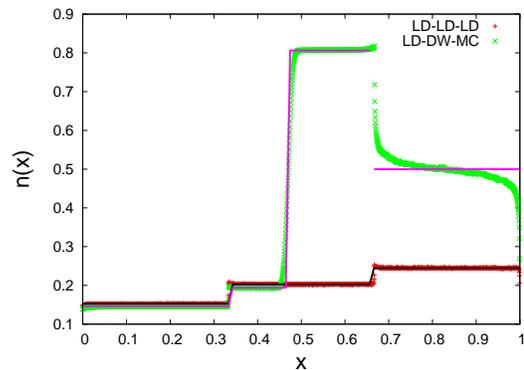}
\caption{Representative plots for (LD-LD-LD, $n=0.2, q=0.7$) and
(LD-DW-MC, $n=0.4, q=0.5$) phases with $p=0.8$ and $N=2100$.
Points and solid lines represent MCS and MFT results, respectively.} \label{3_profile}
\end{figure}

The phase diagram of
the model in the $(n,q)$-plane, parametrized by $p$ is given in
Fig.~(\ref{3_phase}). Its qualitative nature may be discussed in
simple physical terms. The DW-MC region shrinks as $q\rightarrow
0.8$. As $q$ decreases, the DW-MC region becomes wider having its
maximum width as $q\rightarrow 0$. In this limit, CHIII has a
vanishingly small hopping rate and hence the model is nearly
blocked. Considering that the DW phase in CHI/CHII is accompanied
by the MC phase in CHIII, thus CHIII must have a density $1/2$
to be in its MC phase. Notice that the current in the system
in its DW-MC phase is $q/4$ and hence decreases with decreasing $q$.
Therefore, the total number of particles in CHI/CHII, required for
them to reach an LD phase with a system current of $q/4$ and hence,
$n_L$ marking the threshold between the (LD-LD-LD) and (LD-DW-MC)
phases, also comes down with $q$ accordingly.  Using the
particle-hole symmetry, we immediately conclude that the threshold
$n_U$ between the (DW-HD-MC) and (HD-HD-HD) phase increases with {
decreasing} $q$. This explains the increasing width of the (DW-MC)
region for decreasing $q$ in the $(n,q)$ phase diagram. { Taking
density as the order parameter, we note that it changes continuously
across the boundaries between the (LD-LD-LD) and (LD-DW-MC) phases, and
the (DW-HD-MC) and (HD-HD-HD) phases,
 corresponding to second order transitions. This is in contrast to the first
 order transition from the LD to HD phases in an open TASEP. Hence, the usual first
order line in a TASEP is replaced by two second order lines here.}
Our MF results are complemented by extensive MC simulations, which
were realised by using random sequential updates together with
averages over large number of samples. Representative plots of
$n_I(x)$, $n_{II}(x)$ and $n_{III}(x)$ ($N=2100,\,p=0.8$) in the
(LD-LD-LD) and (LD-DW-MC) phases, obtained from both our MFT and MCS
studies, are given in Fig.~\ref{3_profile}. Both figures
(\ref{3_phase}) and (\ref{3_profile}) clearly show that while the
MFT results match qualitatively with the corresponding MCS results,
there are minor quantitative disagreements between them, presumably
due to the correlation effects which are stronger in the closed
model here than for an open TASEP. Notice that there can be
only one LDW in the system, either in CHI or CHII. If it is in CHI
(CHII), then CHII (CHI) will be entirely in its LD (HD) phase,
characterised by uniform densities (neglecting BLs). Nonetheless,
there will however be a jump in the density profile at the junction
B with the jump height being controlled by the difference between
the hopping rates in CHI and CHII respectively. This jump survives
even in the thermodynamic limit. Thus, while in terms of the number
of LDWs, our Model I yields the same results as the FSM
model~\cite{mustansir}, Model I has additional density
discontinuities in its steady states.

\begin{figure}[htb]
\includegraphics[height=5cm]{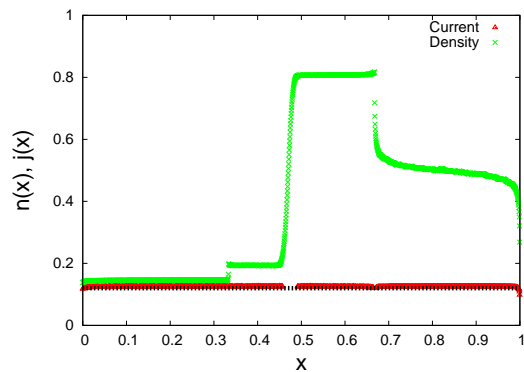}
\caption{Plot showing dependence of the steady state current
in the bulk on the density ($p=0.8$,$q=0.5$,$n=0.4$ and $N=2100$).
The dashed black line represents the MFT result for the current while the points
show MCS results. Since the MCS results for the steady state current is calculated from their mean field expressions,
these are valid only in the bulk, away from the junctions and any discontinuity.} \label{3_current_profile}
\end{figure}

\par
Clearly if we  set $p=1$ in Model I, CHI and CHII become
indistinguishable. In fact, with $p=1$, our Model I effectively
reduces to the two-segment model in Ref.~\cite{mustansir}, such that
the slower part has a length half of the faster part. At this limit,
unsurprisingly our results reduce to those of Ref.~\cite{mustansir}. A
current vs density plot is shown in Fig.~\ref{3_current_profile}, corresponding
to the LD-DW-MC phase in Fig.~\ref{3_profile}.

\section{Model II: Four segment model}\label{four}

\begin{figure}[htb]
\includegraphics[height=5cm]{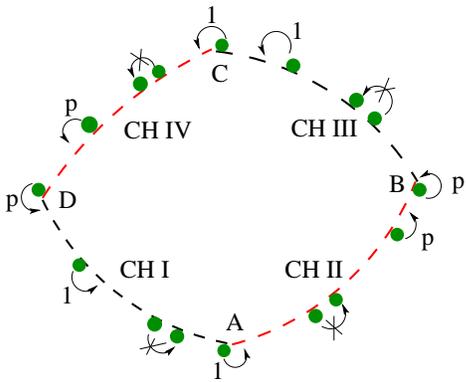}
\caption{Schematic diagram of the 4-sector model. CHI and CHIII (black)
have a hopping rate 1, CHII and CH IV (red) have a hopping rate of $p<1$(see
text).} \label{model2}
\end{figure}

We now extend our Model I to four-segment model (Model II), having
four parts of equal number of sites, but with all unequal hopping
rates along them (say, 1, $p_1,\,p_2,\,p_3$ with $1\geq p_1\geq
p_2\geq p_3$). Extending our arguments and analyses in the previous
Section, it is clear that there are no qualitative changes in the
overall density profiles of Model II, in comparison with Model I:
Either all segments will be in their LD (or HD) phases, or, for
intermediate densities, the segment with the lowest hopping rate
($p_3$) reaches its MC phase, together with the simultaneous
formation of an LDW in any of the other segments with its precise
location being given by the total particle number conservation in
the system. There are, however, surprises, when the different
segments have hopping rates $1,p,1,p$ ($1>p$), respectively (see
Fig.~\ref{model2}). Thus, CHI and CHIII are identical to each other,
where as, CHII and CHIV are mutually identical.


Total particle number conservation and the equality of the current
in the bulk of every segment allow us to calculate the density
profiles in each segment, just as above. Let
$n_I,n_{II},n_{III},n_{IV}$ be the densities of CHI, CHII, CHIIII
and CHIV, respectively. Physically, for very low (high) densities,
all of CHI, CHII, CHIII, CHIV are in their LD (HD) phases, with
$n_I=n_{III}$ and $n_{II}=n_{IV}$ in the bulk. Proceeding as above,
current conservation in the bulk yields
\begin{equation}
n_I(1-n_I)=n_{III}(1-n_{III})=pn_{II}(1-n_{II})=pn_{IV}(1-n_{IV}),\label{4current}
\end{equation}
allowing us to obtain $n_I=n_{III}$ in terms of $n_{II}=n_{IV}$.
Assuming $n_I +n_{II} +n_{III} +n_{IV}=4n$, $n$ being the total
average density, we can thus write,
\begin{equation}
 n_{I} +n_{II}=2n.
\end{equation}
This yields solutions for $n_I$ and $n_{II}$, where,\bea
n_{II}=\frac{-b \pm \sqrt{b^2-4ac}}{2a}, \eea with $a=(p-1)/4,
b=n-\frac{1+p}{4}$ and $c=n/2-n^2$.

As more particles are added to the system, CHII and CHIV should reach
their MC phases (similar to our analysis for Model I above). Then,
\begin{equation}
n_I(1-n_I)=n_{III}(1-n_{III})=p/4,
\end{equation}
since the currents in CHII and CHIV in their MC phases are just
$p/4$. This allows us to solve for $n_I$ and $n_{III}$:
\begin{equation}
n_I=n_{III}=\frac{1 \pm \sqrt{1-p}}{2}
\end{equation}
Thus, for each of $n_I$ in CHI and $n_{III}$ in CHIII,we obtain two
values, one greater and the other less than $1/2$. Following our
physical picture for Model I, thus, an LDW should form connecting
the two solutions for $n_I$ and $n_{III}$ at  points in the bulk of
CHI and CHIII, respectively.

Particle number conservation in the model yields
\begin{equation}\label{4sector_eqn}
\left(\int_0^{1/4}n_I + \int_{1/4}^{1/2}n_{II} +
\int_{1/2}^{3/4}n_{III}+\int_{3/4}^{1}n_{IV}\right)dx = n.
\end{equation}

We now define coordinates $0<x_I<1/4$, $1/4<x_{II}<1/2$,
$1/2<x_{III}<3/4$ and $3/4<x_{IV}<1$ for
CHI, CHII, CHIII and CHIV, respectively
in order to find the location of the LDWs.

Let $x_{Iw}$ and $x_{IIIw}$ be the locations of the LDWs in CHI and
CHIII, respectively. The density profiles $n_I(x_{Iw})$ and
$n_{III}(x_{IIIw})$ may be written as
\begin{equation}
 n_I(x_{Iw})= \alpha + \Theta(x-x_{Iw})(1-\alpha -\beta)
\end{equation}
and
\begin{equation}
 n_{III}(x_{IIIw})= \alpha + \Theta(x-x_{IIIw})(1-\alpha -\beta),
\end{equation}
where, $\alpha=\beta=\frac{1-\sqrt{1-p}}{2}$.

\begin{figure}[htb]
\includegraphics[height=5cm]{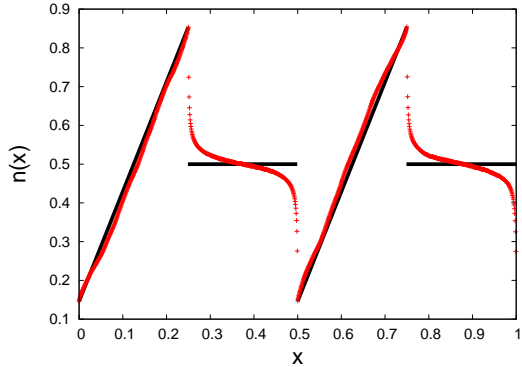}
\caption{Formation of DDWs for $N=1200, n=0.5, p=0.5$.
The solid black lines and the red points represent MFT and MCS results,
respectively. MCS results have been averaged over 10 samples of data. Simulation
results show good agreement with the theory. Each sample was iterated for $2\times10^9$ steps.}\label{4_profile}
\end{figure}

\begin{figure}[htb]
\includegraphics[height=5cm]{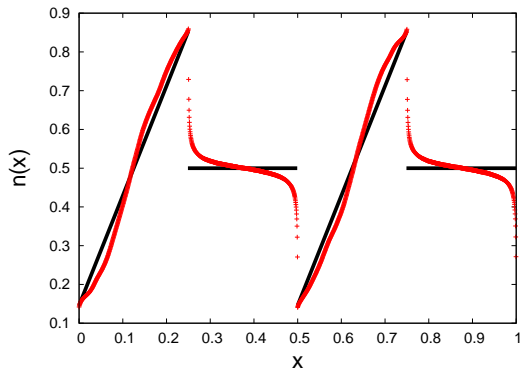}
\caption{Formation of DDWs for $N=2000, n=0.5, p=0.5$.
The solid black lines and the red points represent MFT and MCS results,
respectively. MCS results have been averaged for 32 different samples with each
sample being iterated for $8\times10^9$ steps~\cite{minor_disagreement}.}\label{4_profile_2000}
\end{figure}

\begin{figure}[htb]
\includegraphics[height=5cm]{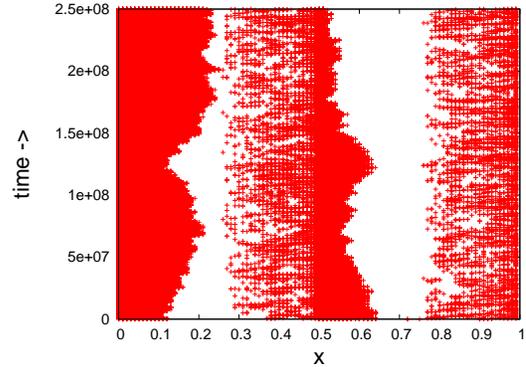}
\caption{Kymograph showing the synchronisation of
 Domain walls in CHI and CHIII. $N=800$,$n=0.5$,$p=0.5$~\cite{kymograph_details}.
 }\label{kymograph_800}
\end{figure}

Substituting for $n_I(x_Iw)$ and $n_{III}(x_{IIIw})$ and noting that
$n_{II}(x_{II})=1/2=n_{IV}(x_{IV})$ in Eq. (\ref{4sector_eqn}) (ignoring the boundary layers) we
find
\begin{equation}
 \frac{1}{4}+\frac{\alpha}{2}+(1-2\alpha)(1-x_{Iw}-x_{IIIw})=n. \label{4sector_con}
\end{equation}
Hence, we have two unknown positions, $x_{Iw}$ and $x_{IIIw}$, of
the LDWs in CHI and CHIII, respectively, but only one equation
(particle number conservation) that relates them. Therefore,
$x_{Iw}$ and $x_{IIIw}$ cannot be determined uniquely, resulting
into {\em delocalised domain walls} (DDW) in each of CHI and CHIII.
The fact that we obtain two DDWs instead of an LDW can be easily
explained from the total particle number conservation. When there is
a single LDW in the system, its position gets fixed by the particle
number conservation. Evidently, particle number conservation can be
maintained equally well for two LDWs by shifting one LDW in one
direction and shifting the second LDW in the reverse direction by
the same amount. Under long time averaging, the system displays an
average of all the (pairwise) LDW positions satisfying total particle
number conservation, yielding two DDWs. Our MFT results on DDWs are corroborated by our MCS studies on Model II; see Fig.~\ref{4_profile} and Fig.~\ref{4_profile_2000}
for DDWs in CHI and CHIII of Model II with two different system sizes.
We present a kymograph (Fig.~\ref{kymograph_800}) depicting synchronisation of the movements of the DDWs in CHI and CHIII.
The synchronised movements of the two DDWs are an outcome of the particle
number conservation as discussed above.

\subsection{DDW profiles and the phase boundaries}
\begin{figure}[htb]
\includegraphics[height=5cm]{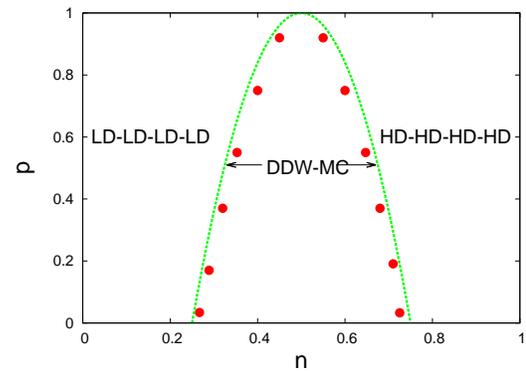}
\caption{Phase diagram of the four segment model. The domain walls
formed are delocalised in contrast to the two and three segment
systems. Points and lines represent MCS and MFT results, respectively.} \label{4_phase}
\end{figure}

While $x_{Iw}$ and $x_{IIIw}$ cannot be determined in MFT uniquely,
their average  profiles may be obtained. Noting that the steady
state densities of CHI and CHIII should be statistically identical,
due to the obvious symmetry between CHI and CHIII, we write $\langle
n_I(x_I)\rangle=\langle n_{III}(x_{III})\rangle$, where $\langle
...\rangle$ represents configuration averages.
 Let us consider a position, $x_0$, in Channel I, where
a DW has formed. Using the symmetry in the system, we can thus say that Channel III will have a DW
at a position, $\frac{1}{2}+x_0$.
So on setting $\left<x_{1w}\right>=x_0$ and $\left<x_{IIIw}\right>=\frac{1}{2}+x_0$, we obtain,
\begin{equation}
 x_0=-\frac{n-\frac{3}{4}+\frac{\alpha}{2}}{2(1-2\alpha)}
\end{equation}
Clearly, the phase boundaries can be obtained by setting $x_0$ equal
to $0$ and $1/4$, respectively. This yields phase boundaries in the
$(n,p)$ plane identical to those obtained for the equal-sized
two-segment system (FSM), described in the beginning. Nonetheless,
the underlying physical pictures are very different. In case of
Model II, the phase boundaries mark the boundaries of the DDW-MC
phases, where as, in case of FSM in Ref.\cite{mustansir}, they refer
to the boundaries of the LDW-MC region. While the topology remains
invariant for the system (see Fig.~\ref{4_phase}), we find that in Model II the LDWs are not
pinned to a particular position, resulting into formation of DDWs.
Evidently, systems with presence of multiple slowest sectors cannot
be reduced to an equivalent two-channel system.  Representative
plots of DDWs in CHI and CHIII, as obtained from both MFT and MCS
studies, are shown in Fig.~\ref{4_profile} and
Fig.~\ref{4_profile_2000}. Good agreement between our MCS are MFT
results are visible.

\section{Possible density profiles for arbitrary number of segments
with unequal hopping rates}\label{general}

Now we generalise our system to arbitrary number  of segments with
unequal hopping rates. For such cases, our results above for three
and four segments may be generalised to obtain information about the
steady state density profiles, which we briefly outline here. For
instance, consider an $n$-segment system with hopping rates
$1>q_2>q_3,...,q_n$, respectively of the segments marked,
$1,2,...,n$. Thus, the $n$th segment  is the slowest segment. Now,
based on our arguments elaborated above, we can conclude that the
$n$th segment should reach its MC phase for intermediate densities.
At the same time, an LDW will be formed
 in any one of the other remaining segments, whose position is controlled by the overall particle
 number density. All the remaining faster segments are to have uniform
 bulk densities. Nevertheless, as in our Model I, there should be
 additional discontinuities in the density profiles at the junctions of the
 different faster segments (i.e, at the junctions between the segments $1,2,...,n-1$).
 Thus, we generally conclude that for any number of finite inhomogeneous segments, all having unequal hopping rates,
 the slowest segment will reach its MC phase for intermediate densities. At the same time, an LDW will be formed
 in any one of the other remaining segments, whose position may be obtained by the overall particle
 number density. Consider now the case where two of the nonconsequtive segments have
 the slowest hopping rates. For concreteness, assume in the system $q_p=q_s$ are the lowest
  hopping rates associated with segments $p$ and $s$, respectively; $p$, $s$ not being neighbours of each other. This would be a generalisation of our Model II above. In this case, again
based on our arguments above, we argue that for intermediate densities two DDWs should
form behind each of the slowest segments ($p$th and $s$th, in the case). In fact,
 if there are several disjoint slowest segments, we expect those many DDWs to form
  behind each of the slowest segments. Calculation of the DDW profiles remains a
   non-trivial task. In the simplest case, the DDW profiles may be obtained by exploiting
   the structural symmetry of the system (as in our Model II). Overall, then, for intermediate
 densities, the system should either have a combination of an MC phase (in the
 slowest segment) and one LDW elsewhere in one of the faster
 segments, or, a combination of several MC phases (when there are more than
 one nonconsecutive slowest segments) and an equal number of DDWs behind each of
 the slowest segments. Of course, for very low or very high densities, all the
 segments should be in either LD or HD phase, regardless of the details of the hopping rates of the individual segments.

\section{Fluctuations  of Domain Walls}\label{scaling}

\subsection{Fluctuations of LDWs}

\begin{figure}[htb]
\includegraphics[height=5cm]{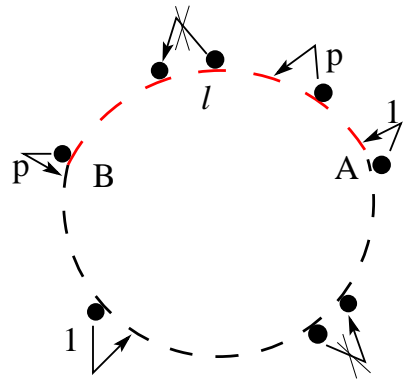}
\caption{Two segment model (FSM) with unequal sizes of
the two segments. The relative length of the defect region is
captured by $l$; $l=1/2$ represents equal sizes of the two segments
(see text).} \label{model_size}
\end{figure}

\begin{figure}[htb]
\includegraphics[height=4.5cm,width=7.5cm]{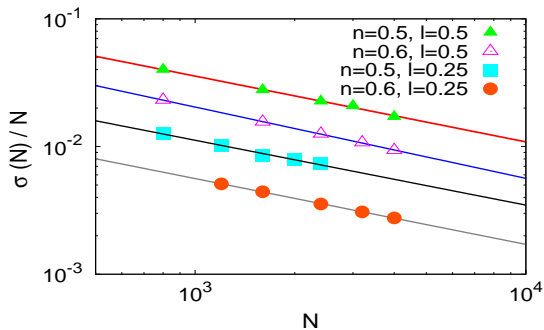}
\caption{Log-log plot of the normalised LDW widths $\sigma(N)/N$
versus $N$: $\sigma(N)/N\sim 1/\sqrt N$ within error bars.(Different
data sets are shifted vertically for better visual
clarity.)}\label{exponent_2}
\end{figure}

In our MF description of Model I, where all fluctuations are
neglected, an LDW is assumed to be static at $x_I^w$. In contrast,
our MCS reveal that an LDW, say in CHI, fluctuates around its mean
position $x^w_I$, which MFT cannot capture. Since there are no
qualitative distinctions between the LDWs in the FSM model of
Ref.~\cite{mustansir} and our Model I, we use the FSM model to
illustrate the fluctuations of the domain walls for simplicity. We
characterise the LDW fluctuations numerically here by means of our
MCS studies. We study the fluctuations for two different sizes of
the defect regions. The size of the defect region relative to the
total system size is characterised by $l$; see Fig.\ref{model_size}.
$l=1/2$ represents the case where the defects cover half of the
system size and $l=0.25$ represents $1/4$th of the system being
covered with defects. The width, $\sigma$ of the distribution of DW
fluctuations can be obtained by fitting the density profile in the
vicinity of the domain wall by the function $(P\cdot
erf[(x-Q)/\sigma]+R)$ \cite{reich}, with the parameters $P, Q, R,
\sigma.$ Our studies reveal normalized width $\sigma (N)/N \sim
1/\sqrt{N}$ (within error bars) for both $n\neq 1/2$ and $n=1/2$.
For $l=1/2,n=0.5$, we have used $N=800,1600,2400,3000,4000$ and for
$l=1/2,n=0.6$, $N=800,1600,2400,3200,4000$. For $l=0.25$ and $n=0.5$, we
have checked with system sizes $800,1200,1600,2000,2400$ and while
$l=0.25$ and $n=0.6$, we have used $N=1200,1600,2400,3200,4000$; see
Fig.~\ref{exponent_2}.
\par
Recall now the studies
in Ref.~\cite{lebodadu} on a ring TASEP with a point bottleneck,
where $\sigma/N$ is shown to scale as $1/\sqrt N$ for $n\neq 1/2$,
where as $\sigma/N\sim 1/N^{2/3}$ for $n=1/2$. The different scaling
law for $n=1/2$ is ascribed to a special symmetry at $n=1/2$ by
which particles and holes, created pairwise at the bottleneck,
impinge upon the DW pairwise~\cite{lebodadu}. In contrast, the FSM
model of Ref.~\cite{mustansir} does not have any special symmetry
for $n=1/2$. This is due to the extended size of CHII (i.e., an
extended defect unlike a point defect in the model of
Ref.~\cite{lebodadu}), for which particles and holes created at the
boundaries between CHI and CHII do not impinge upon the DW for
$n=1/2$ pairwise; in addition, careful observation of the density
profiles obtained in our MCS studies reveal additional BLs in Model
I,
 (see Fig.~\ref{3_profile}), a consequence of correlations borne out of
strict particle conservations. Similar BLs are found in our MCS
studies of the two-segment model (FSM) as well (not shown here). These BLs
appear to survive for all $N$ in our MCS. We expect these BLs should
invalidate the special symmetry at $n=1/2$ of Ref.\cite{lebodadu}
further in the present context.  These factors should be responsible
for our difference in the scaling of $\sigma (N)$ with
Ref.~\cite{lebodadu} for $n=1/2$.
This $1/\sqrt N$ scaling for the width $\sigma(N)/N$ may be
understood heuristically as follows. Due to the stochastic nature of
the microscopic dynamics, the particle current past a given point in
the system is not static even in the steady state; it fluctuates
randomly about its mean value as given by MFT. As a result,  the
number of particles entering (exiting) the segment with an LDW
(i.e., the segment with a higher hopping rate) fluctuates randomly
around their mean values. These fluctuations $\delta N_p$ lead to
the fluctuations in the LDW position, measured by $\sigma$. Let
$N_p$ be the average particle number in that segment. In a 1D system
with exclusion as ours, it is reasonable to assume $\sigma\propto
\delta N_p$. Assuming no long-ranged correlations, we write $\delta
N_p/N_p \sim 1/\sqrt N_p$. Since $N_p$ should scale with $N$ for a
given density and inhomogeneity configuration, we find
$\sigma(N)/N\sim 1/\sqrt N$. In the absence of any special symmetry
at $n=1/2$, we expect the above simple picture should hold good even
for $n=1/2$.

\subsection{Delocalisation transition and fluctuation of DDWs}

As we have seen before, in our Model II with hopping rates $1,p,1,p$
in CHI, CHII, CHIII, CHIV, respectively, the densities in CHI and
CHIII display DDWs for moderate densities, when CHII and CHIV are in
their MC phases. On the other hand, if CHII and CHIV do not have
same hopping rates $p$, there will be only one LDW in the system at
a location governed by the particle number conservation. Therefore,
as the hopping rates in CHII and CHIV approach each other, this LDW
should undergo a delocalisation transition and two DDWs emerge in
CHI and CHIII. Thus, the width of the LDW fluctuation should diverge
in the thermodynamic limit at the delocalisation transition. We
investigate this by MCS studies of our Model II.

\begin{figure}[htb]
\includegraphics[height=5cm]{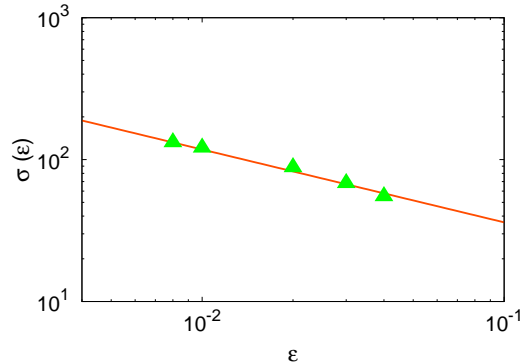}
\caption{Log-log plot for the scaling of the width, $\sigma$ with
$\epsilon$ for the 4-segment system ($N=800, n=0.44$); $\sigma \propto
\epsilon^{-1/2}$. } \label{4_exponent}
\end{figure}

In Fig.~\ref{model2}, assume CHII and CHIV have hopping rates of
$p-\epsilon$ and $p+\epsilon$, respectively. Hopping rates at CHI
and CHIII continues to be unity (as before). Consider $n$ to be such
that CHII be in its MC phase. As $\epsilon \rightarrow 0$, an MC
phase should be formed in CHIV and the LDW should approach a
delocalisation transition, i.e., the width of the LDW is expected to
diverge as $\epsilon\rightarrow 0$. We investigate how the width of
the LDW, $\sigma$ scales with $\epsilon$ by fitting the density
profile in the vicinity of the domain wall by the function $(P\cdot
erf[(x-Q)/\sigma]+R)$, with the parameters $P, Q, R, \sigma.$ (as
above), for a given $N$. MCS studies reveal that, $\sigma \propto
\frac{1}{\epsilon^{1/2}}$ (within error bars; see
Fig.\ref{4_exponent}), diverging as $\epsilon\rightarrow 0$ (or, the
normalized width $\sigma(N)/N$ is a finite value in the limit
$\epsilon\rightarrow 0$ in the thermodynamic limit $N \rightarrow
\infty$). This establishes scaling of the domain wall positions near
the delocalisation transition in our Model II.

A rigorous mathematical formulation of the scaling law $\sigma
\propto \frac{1}{\epsilon^{1/2}}$ may possibly be done by using the
formalism developed in Ref.~\cite{erwin2} appropriately modified for
a closed system. We do not do that here; however, we make
interesting analogies of the delocalisation transitions discussed
above with second order phase transitions in (equilibrium) magnetic
spin systems. Considering the scale of the domain wall fluctuations
as a {\em correlation length} $\xi$ of density fluctuations (since
this sets the scale of the fluctuations of domain wall positions),
we have $\xi\propto \frac{1}{\epsilon^{1/2}}$. This scaling
behaviour is reminiscent of the correlation length $\sim 1/\sqrt
{T-T_c}$ in a magnetic system very close to the critical
temperature $T_c$. Qualitatively, for a magnetic system at $T_c$,
the free energy potential well becomes flat even for large values of
the magnetisation, leading to large magnetisation fluctuations.
Similarly in the present model, the positions of the underlying LDWs
of a DDW envelope can be anywhere within the DDW envelope with equal
probability. Further theoretical investigations going beyond MFT
should be useful in studying this analogy~\cite{warn}.

\section{Conclusion}\label{summ}

In this work, we analyse the generic steady state density profiles
in TASEP on a ring with extended inhomogeneities. We find for
moderate average densities, nontrivial spatial dependences for the
density profiles in the steady states, including LDW, along with
second order transitions, and localisation-delocalisation transition
leading to DDWs. The underlying inhomogeneity configuration together
with the overall particle number density controls the macroscopic
density profiles. For instance, we demonstrate that for very low
(high) densities, the system is in homogeneous LD (HD) phases,
regardless of the inhomogeneity configuration (i.e., for both Model
I and Model II above). For moderate densities, the underlying
inhomogeneity configurations become crucial. We find that in our
Model I, having three segments with unequal hopping rates, the
overall behavior is qualitatively similar to that of the FSM model
of Ref.~\cite{mustansir}, having two segments of unequal hopping
rates. Both Model I and FSM display one LDW for moderate densities,
the location of which is determined essentially by the particle
number conservation. When the number of segments increases to four
(Model II), with all unequal hopping rates, again there is
generically only one LDW for intermediate densities, similar to our
Model I. However, there are now additional density discontinuities
at the  junctions between the faster segments. Surprisingly, in
Model II for specially constructed inhomogeneity configurations with
hopping rates $1, p, 1, p;\, p<1$, the macroscopic density profiles
are significantly different in having two DDWs, instead of one LDW.
We argue that this is due to the existence of more than one (two
here) slowest segments in Model II. In general, our MFT and MCS
results match qualitatively for both Model I and Model II, in
general.  While the instantaneous DDW positions cannot be determined
cannot be determined from MFT, we discuss how their long-time
averaged density profiles may be constructed by using MFT. We use
our detailed results from Model I and Model II to infer about the
possible steady state density profiles and argue that depending upon
whether the system has one or more than one slowest segments, either
the system should show an LDW or a number of DDWs (equal to the
number of the slowest segments). Notice that the
current-density relationship in a given segment is formally
continued to be given by the same expression as for an isolated open
TASEP with an arbitrary hopping rate. Equality of the steady state
currents in all the segments ensures that in the slowest segment(s),
the current can reach its maximum value corresponding to the MC
phase(s). Thus the form of the current versus density plots for the
slowest segments should be identical in nature to that for an open
TASEP. In contrast, the maximum currents in the faster segments will
be limited by the extended defects (the slowest segments), as these
segments will not reach their MC phases. Thus, the currents in these
faster segments will be less than their values in their putative MC
phases had they been isolated open TASEPs. Hence, in the
corresponding current-density plots, the maximum value of the
current will be truncated to a value controlled by the MC phase
currents in the slowest segments, see for
example~\cite{mustansir},~\cite{stinch1}. Lastly, we study the
fluctuations of LDW and DDWs numerically. We show that the relative
width of an LDW scales as $1/\sqrt N$, such that in the
thermodynamic limit $N\rightarrow\infty $, an LDW becomes fully
sharp. We also numerically illustrate the delocalisation transitions
of an LDW in our Model II, when the hopping rates of two of the
slower segments approach each other. Assuming hopping rates of the
four segments in Model II as $1,p+\epsilon,1,p-\epsilon$, two DDWs
are formed, with their spans diverging in the thermodynamic limit
$N\rightarrow\infty$ for $\epsilon=0$. We numerically establish that
as $\epsilon\rightarrow 0$, the width, $\sigma\propto
1/\sqrt\epsilon$~\cite{div}. Overall, at the broader level our
work reveals how the mutual interplay between the particle number
conservation and the quenched disorder configurations yield
different types of density profiles including LDW and DDWs.

 Our model serves as a simple starting point to study the effects of an arbitrary number of extended inhomogeneities
 on asymmetric exclusion processes in a closed system.   Additional biologically relevant details
may be incorporated in our model, e.g., the detailed features of the
model discussed in Ref.~\cite{dc}. On the theoretical side, how
nonconservation of particles may affect the density profiles
obtained here should be a worthwhile question to study. Due to the
minimalist nature of the model, we do not expect immediate
quantitative agreement with specific experimental results.
Nevertheless, qualitative features of our results may be tested in
ribosome density mapping~\cite{arava} or ribosome
profiling~\cite{ingolia} experiments on model mRNA loops with
clusters of slow codons. In particular, it may be noted that
arguments those yield the existence of LDW or DDWs in our models for
intermediate densities are independent of the precise lengths of the
extended defects. Thus, while we have assumed equal lengths for the
segments in both Model I and Model II for simplicity, we expect the
qualitative features of LDW and DDW should survive even for short
segments of several extended defects. This may be of particular
interests in the context of mRNA loops with slow codons, where the
latter generally occur in several short segments.  In the context of
vehicular traffic flows, our model shows under what condition there
may be crowding of vehicles behind  extended bottlenecks in a closed
network of roads. Therefore, it is expected that in a closed network
of narrow roads with no possibility of overtaking, one would
generically observe either moving jams (DDWs) or static jams (LDW),
depending upon the geometry of the defects. In addition, possible
experimental realisations on the unidirectional restricted motion of
interacting particles in narrow micropores (with the mutual passage
excluded), e.g., in closed channels with circular geometry with
extended bottlenecks~\cite{beching} should make our model an
intriguing system to study.

\paragraph{Acknowledgement:-} We thank the anonymous Referees for their constructive suggestions in course of the review process. One of the authors (AB) wishes to thank  the
Max-Planck-Gesellschaft (Germany) and the Department of Science and
Technology/Indo-German Science and Technology Centre (India) for
partial financial support through the Partner Group programme
(2009).


\begin{thebibliography}{99}
\bibitem{kymograph_details} The mesoscopic length, say $r$, used to
draw a kymograph, i.e., the length over which the spatial averaging
is done should be much smaller than N. We have used $r=6$.
\bibitem{warn} It will be interesting to see how strictly the
analogy between delocalisation transitions here and phase
transitions in equilibrium magnetic systems hold due to the inherent
nonequilibrium nature of the dynamics in the present model.
\bibitem{div} For a finite system, this divergence will be cut off
by the finiteness of the system. This is analogous to rounding of
the critical divergenges at the critical temperature in finite
systems. In the theoretical $N\rightarrow \infty$ limit, the scaling
law $\sigma\propto 1/\sqrt\epsilon$ should be observed without any
rounding effect in the theromodynamic limit.
\end{thebibliography}
\end{document}